\documentclass[thmsa,12pt]{article}
\usepackage{amsmath}

\input{tcilatex}

\begin{document}

\setcounter{page}{0} \topmargin0pt \oddsidemargin5mm \renewcommand{%
\thefootnote}{\fnsymbol{footnote}} \newpage \setcounter{page}{0} 
\begin{titlepage}
\begin{flushright}
YITP-SB-00-71 \\
\end{flushright}
\vspace{0.5cm}
\begin{center}
{\Large {\bf Colours associated to non simply-laced\\
 Lie algebras and exact S-matrices} }

\vspace{0.8cm}
{\large Christian Korff }

\vspace{0.5cm}
{\em C.N. Yang Institute for Theoretical Physics\\
State University of New York at Stony Brook\\ 
Stony Brook, N.Y. 11794-3840}
\end{center}
\vspace{0.2cm}
 
\renewcommand{\thefootnote}{\arabic{footnote}}
\setcounter{footnote}{0}

\begin{abstract}
A new set of exact scattering matrices in 1+1 dimensions is proposed by solving the bootstrap equations. 
Extending earlier constructions of colour valued scattering matrices this new set has its colour structure 
associated to non simply-laced Lie algebras. This in particular leads to a coupling of different affine 
Toda models whose fusing structure has to be matched in a suitable manner. The definition of the new S-matrices 
is motivated by the semi-classical particle spectrum of the non simply-laced Homogeneous 
Sine-Gordon (HSG) models, which are integrable perturbations of  WZNW cosets. In particular, the 
S-matrices of the simply-laced HSG models are recovered as a special case.
\end{abstract}
\vfill{ \hspace*{-9mm}
\begin{tabular}{l}
\rule{6 cm}{0.05 mm}\\
korff@insti.physics.sunysb.edu \\
\end{tabular}}
\end{titlepage}
\newpage

\section{Introduction}

In the context of 1+1 dimensional integrable quantum field theories the idea
of analytic continuation allows for the explicit construction of exact
scattering amplitudes by means of the bootstrap approach \cite{boot}. In
integrable field theories each scattering process can be decomposed into
two-particle processes reducing the problem of calculating the S-matrix to
determining the two-particle amplitude $S_{AB}(\theta )$, where $A,B$ label
the particle types and $\theta $ is the rapidity variable. The analytic
continuation of $S_{AB}$ is subject to the following set of functional
equations, 
\begin{eqnarray}
S_{AB}(\theta )S_{BA}(-\theta ) &=&1  \label{B1} \\
S_{AB}(\theta -i\pi )S_{A\bar{B}}(\theta ) &=&1  \label{B2} \\
S_{DA}(\theta )S_{DB}(\theta +iu_{AB}^{C})S_{DC}(\theta
+iu_{AB}^{C}+iu_{BC}^{A}) &=&1\;.  \label{B3}
\end{eqnarray}
Equations (\ref{B1}) and (\ref{B2}) reflect the physical constraints of
unitarity and crossing symmetry, respectively. Equation (\ref{B3}) is the
bootstrap requirement associated with the fusing process $A+B\rightarrow 
\bar{C}$ and plays the most important role in the construction\footnote{%
The Yang-Baxter \cite{YB}, which in general also arises in this context,
will be trivially fulfilled, since we are going to consider only theories
where the S-matrices are assumed to be diagonal.}, since it incorporates the
bound state structure of the integrable quantum field theory via the fusing
angles $u_{AB}^{C}$. Which fusing processes occur is usually inferred from
the classical Lagrangian of the theory either by means of perturbation
theory or a semi-classical analysis. In this manner the solution of the
bootstrap equations is linked to a specific field theory identified by a
concrete Lagrangian.\medskip

However, one might look for solutions of the functional equations
independent whether or not a classical Lagrangian formulation of a field
theory is given. Indeed, one of the remarkable messages of the bootstrap
approach is that every consistent solution to equations (\ref{B1})-(\ref{B3}%
) can be interpreted as a two-particle scattering amplitude defining
implicitly an integrable quantum field theory. This point of view motivated
the construction of a whole set of new factorizable scattering matrices with
colour values in \cite{colour} by extending S-matrices proposed earlier \cite
{HSGS} in the context of the so-called simply-laced Homogeneous Sine-Gordon
(HSG) models \cite{HSG} to a much larger class. In the present work the
construction scheme outlined in \cite{colour} will be generalized to involve
also non simply-laced Lie algebras for the colour values by choosing the
semi-classical particle spectrum \cite{HSGsol} of the so-called \emph{non
simply-laced} HSG models as input data for the bootstrap (\ref{B3}). The
resulting factorizable S-matrices might be interpreted as possible
candidates for these integrable quantum field theories. As a preparatory
step for the construction it is briefly recalled how the Lie algebraic
structures enter in the S-matrix construction.\medskip

The Lie algebraic methods used in \cite{colour} to provide consistent
solutions to the bootstrap equations originate in affine Toda field theory 
\cite{ATFT}. The key feature exploited is the splitting of the $ADE$ affine
Toda S-matrix \cite{TodaS} into a minimal part $\mathcal{S}^{g^{\prime }}$
containing all the physical poles and a CDD-factor \cite{CDD} $\mathcal{F}%
^{g^{\prime }}$ displaying the coupling dependence. Here the simple Lie
algebra $g^{\prime }=ADE$ fixes the affine Toda model in question. As
explained in \cite{colour} this model can be 'multiplied' by labeling
particles through two quantum numbers $A=(a,i)$ each corresponding to a
vertex in the Dynkin diagrams of a \emph{pair} of simple \emph{simply-laced}
Lie algebras $g^{\prime }|g$. As already pointed out the first algebra $%
g^{\prime }$ fixes the scattering amplitude from affine Toda field theory
(ATFT), while the second yields the so-called colour degree of freedom
governing the interaction type. Particles of the same colour $i$ are chosen
to interact via the minimal affine Toda S-matrix $S_{ab}^{ii}(\theta ):=%
\mathcal{S}_{ab}^{g^{\prime }}(\theta ),$ while particles of different
colours $i,j$ scatter via the CDD-factor provided the corresponding vertices
in the Dynkin diagram of $g$ are linked to each other, 
\begin{equation}
S_{ab}^{ij}(\theta ):=\left\{ 
\begin{array}{cc}
1\;, & A_{ij}=0 \\ 
\eta _{ab}^{ij}\mathcal{F}_{ab}^{g^{\prime }}(\theta ,B=1)^{\frac{1}{2}}\neq
1\;, & A_{ij}<0
\end{array}
\right. \;.  \label{Def}
\end{equation}
Here $A$ is the Cartan matrix associated with the colour Lie algebra $g$ and
the square root has been taken of the CDD-factor $\mathcal{F}^{g^{\prime }}$
at effective coupling $B=1$. The factor $\eta _{ab}^{ij}=\bar{\eta}%
_{ba}^{ji} $ in front is a constant of modulus one crucial for satisfying
the bootstrap requirements and leading to parity violation, $%
S_{ab}^{ij}(\theta )\neq S_{ba}^{ji}(\theta )$ (see \cite{HSGS,colour} for
details). Note that in order to avoid 'bulky' expressions the notation used
in \cite{colour} has been adopted denoting the scattering amplitude $S_{AB}$
with $A=(a,i)$, $B=(b,j)$ by $S_{ab}^{ij}$. \medskip

\noindent Besides yielding numerous new solutions it was shown that the
above construction (\ref{Def}) also recovers several known cases as specific
subsets, one being the scaling or minimal affine Toda models \cite
{Zper,TodaS} when choosing $g^{\prime }|su(2)$ and the other the already
mentioned class of simply-laced HSG theories when selecting $su(k)|g$. The
common feature of these theories is their interpretation as integrable
perturbations of parafermionic conformal field theories \cite{para,Gepner}.
The latter are obtained by the GKO coset construction \cite{GKO} from WZNW
models \cite{Witten}. For example the minimal affine Toda or scaling models
are related to cosets of the form $g_{1}^{\prime }\oplus g_{1}^{\prime
}/g_{2}^{\prime },$ while the HSG theories are linked to $g_{k}/u(1)^{\times 
\limfunc{rank}g}$. Here the lower index refers to the so called \emph{level}
labeling the representation of the affine extension of $g,g^{\prime }$. This
matching between exact S-matrices constructed via the bootstrap approach and
conformal field theories can be achieved by several methods.

For instance, the thermodynamic Bethe ansatz (TBA) \cite{TBA0,TBAZ} allows
to calculate the effective central charge of the underlying conformal models
from the two-particle scattering amplitude. For the cases mentioned this
analysis has been carried out in \cite{TBAZ}\ and \cite{CFKM}. The general
case of all simply-laced colour valued S-matrices was discussed in \cite
{colour}\ and the universal formula found for the effective central charge
matches with the one obtained by Dunne et al. \cite{Dunne} when generalizing
the discussion of parafermionic CFT's in \cite{Gepner}. An alternative
approach which enables one to extract information about the underlying UV
conformal model is the form factor program \cite{FF}. In \cite{CFK,CF} it
has been shown for a series of simply-laced HSG models that besides the
central charge even the conformal dimensions and the local operator content
can be extracted. \medskip

\noindent Looking at the conformal cosets mentioned above it is natural to
ask about possible extensions when $g^{\prime },g$ are chosen to be \emph{%
non simply-laced} simple Lie algebras, i.e. of $BCFG$ type. The construction
of S-matrices involving non simply-laced algebras usually turns out to be
more complicated. For instance, selecting $g^{\prime }$ to be non
simply-laced the mentioned separation property of the affine Toda S-matrix
is spoiled \cite{Delius&al} due to a coupling dependent flow between two
dual algebras (see e.g. \cite{FKS2,thesis} and references therein). Hence,
the construction scheme of \cite{colour}\ breaks down. Nonetheless, in this
work it will be demonstrated how the construction can at least be extended
to non simply-laced colour algebras $g.$ It will turn out that the non
simply-laced structure of $g$ can be accommodated by coupling different ATFT
models to each other, i.e. the choice of $g^{\prime }$ will vary in
dependence on the Lie algebraic data of $g$. \medskip

\noindent The article is organized as follows. Section 2 is concerned with
constructing the set of new S-matrices with non simply-laced colour
structure and starts out with defining the asymptotic spectrum of particles.
Motivated by the semi-classical analysis of non simply-laced HSG models \cite
{HSGsol} the different length of the simple roots of $g=BCFG$ will be taken
into account by choosing different particle numbers for each colour value $%
i=1,...,\limfunc{rank}g$. In fact, this construction aims at an extension of
the simply-laced HSG S-matrices, whence the algebra $g^{\prime }$ will be
chosen to be $su(k_{i})$ where the value of $k_{i}>1$ will depend on the
colour as explained below. In the next step closed formulas are provided for
the new set of S-matrices covering all choices of $g$ including the
simply-laced algebras. Using the same techniques as in \cite{colour} the new
S-matrices are shown to be consistent solutions of the functional equations (%
\ref{B1}), (\ref{B2}) and (\ref{B3}). To demonstrate the working of the
general formulas the case $g=B_{2}$ or $so(5)$ is presented as an example.
Section 3 states the conclusions.

\section{Colours from non simply-laced Lie algebras}

In the following let $g$ denote a simple Lie algebra, $h^{\vee }$ its dual
Coxeter number and $\{\alpha _{i}\}_{i=1}^{\limfunc{rank}g}$ a set of simple
roots. Normalizing the length of the long roots to be $\alpha _{\text{long}%
}^{2}=2$ it will turn out to be convenient for the subsequent calculations
to define the integers 
\begin{equation}
t_{i}:=\frac{2}{\alpha _{i}^{2}}\in \{1,2,3\}
\end{equation}
which symmetrize the Cartan matrix $A$ associated to $g$, 
\begin{equation}
A_{ij}t_{j}=A_{ji}t_{i}\quad \text{with}\quad A_{ij}=t_{i}\left\langle
\alpha _{i},\alpha _{j}\right\rangle \;.
\end{equation}
Motivated by the semi-classical particle spectrum found in \cite{HSGsol}
assign to each simple root $\alpha _{i}$ a tower of stable particles whose
mass ratios are determined by 
\begin{equation}
M_{a}^{i}=m_{i}\sin \frac{\pi a}{k_{i}}\;,\quad \quad i=1,...,\limfunc{rank}%
g,\quad \quad a=1,...,k_{i}-1,  \label{mass}
\end{equation}
where $k_{i}:=t_{i}k$ and $k>1$ is the so-called \emph{level}. Thus,
analogous to the discussion in \cite{colour} the stable particles are
labelled by a pair $(a,i)$ of quantum numbers and the choice (\ref{mass})
links the structure of the mass spectrum for fixed $i$ to the $su(k_{i})$
affine Toda model. The $\limfunc{rank}g$ constants $m_{i}$ are left
undetermined and might be all the same or different\footnote{%
Notice that it is always assumed that the quantum particles are
distinguishable by some unspecified quantum charge justifying the ansatz of
diagonal S-matrices.}. Since the mass spectrum resembles the one from ATFT
it also inherits the corresponding fusing structure determined by the fusing
angles \cite{TodaS} 
\begin{equation}
u_{ab}^{c}=\left\{ 
\begin{array}{cc}
\frac{\pi }{k_{i}}(a+b), & a+b+c=k_{i} \\ 
2\pi -\frac{\pi }{k_{i}}(a+b), & a+b+c=2k_{i}
\end{array}
\right. \;.
\end{equation}
Here it is understood that whenever the above conditions on the particle
indices $a,b,c$ are not satisfied the fusing process $a+b\rightarrow \bar{c}$
is not present in the theory. As discussed in the introduction the
corresponding affine Toda scattering amplitude can be expressed as a product
of a so-called minimal and a CDD-factor \cite{TodaS}, 
\begin{equation}
\text{ATFT:\quad \quad }\mathcal{S}_{ab}^{su(k_{i})}(\theta ,B)=\mathcal{S}%
_{ab}^{su(k_{i})}(\theta )\mathcal{F}_{ab}^{su(k_{i})}(\theta ,B),
\end{equation}
each of which satisfies the functional equations (\ref{B1})-(\ref{B3})
separately.

In order to proceed as closely as possible to the former construction the
particles for fixed quantum number $i$ are assumed to interact via the
minimal scattering matrix of $su(k_{i})$ ATFT, 
\begin{equation}
S_{ab}^{ii}(\theta )=\mathcal{S}_{ab}^{su(k_{i})}(\theta )\;.  \label{def0}
\end{equation}
In contrast to the simply-laced case one has now \emph{different} models of
ATFT coupled to each other. Thus, when looking for a CDD-interaction between
particles of different colours $i,j$, which is similar to the one explained
in the introduction, the first problem one encounters is that the quantum
numbers $a,b$ may now run over different ranges, namely $a=1,...,k_{i}-1$
and $b=1,...,k_{j}-1$. This obviously excludes the possibility of taking
always the same CDD-factor as in the simply-laced case. Secondly, the
CDD-factor has now to comply with both the fusing structure of $su(k_{i})$
and $su(k_{j})$ ATFT. These two problems can be resolved by choosing the
CDD-factor belonging to the affine Toda model associated with the shorter
root and by a suitable identification of the particles in the different
copies.\medskip

Let us assume that $t_{i}>1$ and $t_{j}=1$, which is the only problematic
case coupling two different affine Toda models to each other. Because of its
particular simple form the fusing condition of $su(k)$ ATFT \cite{TodaS} 
\begin{equation}
su(k):\quad \quad a+b\rightarrow \bar{c}\quad \Longleftrightarrow \quad
a+b+c=k\text{ or }2k
\end{equation}
can be easily translated into a fusing rule of $su(k_{i})$ ATFT by
multiplication with the constant $t_{i}$ and by identifying the particles $a$
and $t_{i}a$ in the $su(k)$ and $su(k_{i})$ theory, 
\begin{equation}
su(k_{i}):\quad \quad t_{i}a+t_{i}b\rightarrow t_{i}\bar{c}\quad
\Longleftrightarrow \quad t_{i}a+t_{i}b+t_{i}c=k_{i}\text{ or }2k_{i}
\end{equation}
Moreover, also the corresponding fusing angles of $su(k)$ and $su(k_{i})$
coincide under the above prescription, 
\begin{equation}
\left. u_{ab}^{c}\right| _{su(k)}=\left. u_{t_{i}a,t_{i}b}^{t_{i}c}\right|
_{su(k_{i})}
\end{equation}
Therefore, the $su(k_{i})$\ affine Toda CDD-factor $\mathcal{F}^{su(k_{i})}$
satisfies the bootstrap equation (\ref{B3}) related to $su(k)$, 
\begin{equation}
\mathcal{F}_{d,t_{i}a}^{su(k_{i})}(\theta )\mathcal{F}%
_{d,t_{i}b}^{su(k_{i})}(\theta +iu_{ab}^{c})\mathcal{F}%
_{d,t_{i}c}^{su(k_{i})}(\theta +iu_{ab}^{c}+iu_{bc}^{a})=1\;.  \label{bootA}
\end{equation}
Thus, when analogously to (\ref{Def}) we set 
\begin{equation}
S_{ab}^{ij}(\theta )=\left\{ 
\begin{array}{cc}
1\;, & A_{ij}=0 \\ 
\eta _{ab}^{ij}\mathcal{F}_{a,t_{i}b}^{su(k_{i})}(\theta ,B=1)^{\frac{1}{2}%
}\neq 1\;, & A_{ij}<0
\end{array}
\right.  \label{def}
\end{equation}
the matrix element $S_{ab}^{ij}(\theta )$ satisfies the necessary bootstrap
equations of both algebras $su(k_{i})$ and $su(k)$ provided that 
\begin{equation}
\eta _{ab}^{ij}=\exp i\pi \varepsilon _{ij}\left( A_{su(k_{i})}^{-1}\right)
_{\bar{a},\,t_{i}b}^{{}}\;
\end{equation}
with $\varepsilon _{ij}=-\varepsilon _{ji}$ being the antisymmetric tensor, $%
A_{su(k_{i})}$ the Cartan matrix of $su(k_{i})$ and $\bar{a}=k_{i}-a$ the
antiparticle of $a$. The insertion of the above phase factor becomes
necessary for the same reason as in the simply-laced case \cite{colour}.
(Note that the charge conjugation in the particle index is missing there).
Having taken the square root of the CDD-factor in (\ref{def}) it still
satisfies the bootstrap equation (\ref{bootA}) with the possible exception
of an overall minus sign. It is exactly this possible sign change which is
compensated by the above phase factor and enforces the breaking of parity
invariance \cite{HSGS,colour}. This can be explicitly checked when
expressing the S-matrix elements $S_{ab}^{ij}$ in terms of hyperbolic
functions.

Adopting a similar notation as in \cite{colour} the following blocks of
meromorphic functions will allow for a compact definition of the S-matrices, 
\begin{equation}
\left[ x\right] _{\theta ,ij}=e^{\frac{i\pi x\,\varepsilon _{ij}}{k_{ij}}%
}\left( \frac{\sinh \tfrac{1}{2}(\theta +i\pi \frac{x-1+B_{ij}}{k_{ij}}%
)\sinh \tfrac{1}{2}(\theta +i\pi \frac{x+1-B_{ij}}{k_{ij}})}{\sinh \tfrac{1}{%
2}(\theta -i\pi \frac{x-1+B_{ij}}{k_{ij}})\sinh \tfrac{1}{2}(\theta -i\pi 
\frac{x+1-B_{ij}}{k_{ij}})}\right) ^{\frac{1}{2}}\,\,  \label{bl}
\end{equation}
with $\varepsilon _{ij}$ being the anti-symmetric tensor already defined
above, $t_{ij}=\max (t_{i},t_{j}),k_{ij}=t_{ij}k$ and $%
B_{ij}=I_{ij}t_{j}/t_{ij}$ the symmetrized incidence matrix $I=2-A$. This
block can be easily seen to have the obvious properties 
\begin{equation}
\left[ x\right] _{\theta ,ij}\left[ x\right] _{-\theta ,ji}=1\qquad \text{%
and\qquad }\left[ k_{ij}-x\right] _{\theta ,ij}=\left[ x\right] _{i\pi
-\theta ,ji}\quad \text{for\quad }B_{ij}=1.  \label{prop}
\end{equation}
In the second equality it is implied that one first takes the square root
and thereafter performs the shifts in the arguments. Note further that the
order of the colour values is relevant, indicating the absence of parity
invariance. We are now prepared to define the S-matrix in a closed formula
including the two special cases of the same (\ref{def0}) and different
colour values (\ref{def}) discussed before. In terms of meromorphic
functions it reads\footnote{%
The expression (\ref{S}) in terms of blocks of meromorphic functions can
also be written by use of Coxeter geometry similar as in \cite{colour}.
However, since we are dealing here only with $su(k_{i})$ ATFT (\ref{S}) is
sufficient to check the bootstrap properties.}, 
\begin{equation}
S_{ab}^{ij}(\theta )=\prod_{\substack{ t_{ij}|a/t_{i}-b/t_{j}|+1 \\ \text{%
step}2}}^{t_{ij}(a/t_{i}+b/t_{j})-1}\left[ x\right] _{\theta
,ij}^{A_{ij}t_{j}/t_{ij}}\;.  \label{S}
\end{equation}
Besides (\ref{S}) there is another representation of the S-matrix which is
usually of advantage when discussing the thermodynamic Bethe ansatz or the
form factor approach to correlation functions \cite{FF}, 
\begin{equation}
S_{ab}^{ij}(\theta )=\eta _{ab}^{ij}\exp \int \frac{dt}{t}\,e^{-it\theta
}\left( 2\cosh \frac{\pi t}{k_{i}}\,\delta _{ij}-I_{ij}t_{j}/t_{ij}\right)
\left( 2\cosh \frac{\pi t}{k_{ij}}-I_{su(k_{ij})}\right) _{\frac{t_{ij}}{%
t_{i}}a,\,\frac{t_{ij}}{t_{j}}b}^{-1}\,.  \label{S1}
\end{equation}
with the phase factor in front equal to 
\begin{equation}
\eta _{ab}^{ij}:=\exp i\pi \varepsilon _{ij}\left(
A_{su(k_{ij})}^{-1}\right) _{\frac{t_{ij}}{t_{i}}\bar{a},\,\frac{t_{ij}}{%
t_{j}}b}\;.  \label{phase}
\end{equation}
By a straightforward calculation similar to those in the simply-laced case,
one can now verify by exploiting (\ref{prop}) that the above S-matrix
satisfies the correct bootstrap properties \cite{colour,thesis}. The
complications which arise due to the different fusing structures for
different colour values have been already discussed above and are taken into
account by a suitable identification of the particles in the $su(k_{i})$ and 
$su(k_{j})$ affine Toda model.

Note that the above expressions reduce to the already derived S-matrix of
simply-laced HSG models by setting $t_{ij}=t_{i}=1$ in accordance with the
above definitions. For instance, formula (\ref{S1}) then simplifies to 
\begin{equation*}
S_{ab}^{ij}(\theta )=e^{i\pi \varepsilon _{ij}[A_{su(k)}^{-1}]_{\bar{a}%
b}}\exp \int \frac{dt}{t}\,e^{-it\theta }\left( 2\cosh \frac{\pi t}{k}%
-I\right) _{ij}\left( 2\cosh \frac{\pi t}{k}-I_{su(k)}\right) _{ab}^{-1}
\end{equation*}
which coincides with earlier expressions found in \cite{HSGS,CFKM,colour}.

\subsection{Example $g=B_{2}$ or $so(5)$}

In order to demonstrate the working of the general formulas this section
presents a specific example for the choices $g=B_{2}$ and $k=3$. The
convention is chosen to label by $i=1$ the long root and by $i=2$ the short
root, i.e. $t_{1}=1,t_{2}=2$. Thus, two copies of the $su(k_{1}=3)$ and $%
su(k_{2}=6)$ affine Toda model will be coupled to each other. The $su(3)\,$%
copy contains two particle types whose scattering matrix is given by 
\begin{equation}
S_{ab}^{11}(\theta )=\left( 
\begin{array}{cc}
\lbrack 1]_{\theta ,11}^{2} & [2]_{\theta ,11}^{2} \\ 
\lbrack 2]_{\theta ,11}^{2} & [1]_{\theta ,11}^{2}
\end{array}
\right) \;.
\end{equation}
Instead the $su(6)$ copy contains five particles whose scattering amplitudes
(\ref{S}) in terms of the meromorphic blocks (\ref{bl}) read 
\begin{equation*}
S_{ab}^{22}(\theta )=\left( 
\begin{array}{ccccc}
\lbrack 1]_{\theta ,22}^{2} & [2]_{\theta ,22}^{2} & [3]_{\theta ,22}^{2} & 
[4]_{\theta ,22}^{2} & [5]_{\theta ,22}^{2} \\ 
\lbrack 2]_{\theta ,22}^{2} & [1]_{\theta ,22}^{2}[3]_{\theta ,22}^{2} & 
[2]_{\theta ,22}^{2}[4]_{\theta ,22}^{2} & [3]_{\theta ,22}^{2}[5]_{\theta
,22}^{2} & [4]_{\theta ,22}^{2} \\ 
\lbrack 3]_{\theta ,22}^{2} & [2]_{\theta ,22}^{2}[4]_{\theta ,22}^{2} & 
[1]_{\theta ,22}^{2}[3]_{\theta ,22}^{2}[5]_{\theta ,22}^{2} & [2]_{\theta
,22}^{2}[4]_{\theta ,22}^{2} & [3]_{\theta ,22}^{2} \\ 
\lbrack 4]_{\theta ,22}^{2} & [3]_{\theta ,22}^{2}[5]_{\theta ,22}^{2} & 
[2]_{\theta ,22}^{2}[4]_{\theta ,22}^{2} & [1]_{\theta ,22}^{2}[3]_{\theta
,22}^{2} & [2]_{\theta ,22}^{2} \\ 
\lbrack 5]_{\theta ,22}^{2} & [4]_{\theta ,22}^{2} & [3]_{\theta ,22}^{2} & 
[2]_{\theta ,22}^{2} & [1]_{\theta ,22}^{2}
\end{array}
\right) \;.
\end{equation*}
Note that all blocks appear squared compensating the square root in the
definition (\ref{bl}), whence all elements are meromorphic. Identifying the
particle $a=1,2$ in the $su(3)$ copy with the particles $a=2,4$ in the $su(6)
$ copy yields then the S-matrix elements 
\begin{equation*}
S_{ab}^{12}(\theta )=S_{\bar{b}a}^{21}(i\pi -\theta )=\left( 
\begin{array}{cc}
\lbrack 2]_{\theta ,12}^{-1} & [4]_{\theta ,12}^{-1} \\ 
\lbrack 1]_{\theta ,12}^{-1}[3]_{\theta ,12}^{-1} & [3]_{\theta
,12}^{-1}[5]_{\theta ,12}^{-1} \\ 
\lbrack 2]_{\theta ,12}^{-1}[4]_{\theta ,12}^{-1} & [2]_{\theta
,12}^{-1}[4]_{\theta ,12}^{-1} \\ 
\lbrack 3]_{\theta ,12}^{-1}[5]_{\theta ,12}^{-1} & [1]_{\theta
,12}^{-1}[3]_{\theta ,12}^{-1} \\ 
\lbrack 4]_{\theta ,12}^{-1} & [2]_{\theta ,12}^{-1}
\end{array}
\right) \;.
\end{equation*}
Again the S matrix elements contain only meromorphic functions, since here
the colours are different, whence $B_{12}=B_{21}=1$ in (\ref{bl}). Note that
there now appear additional phase factors in the blocks (\ref{bl}) which
break parity invariance. By a straightforward calculation one now verifies
the bootstrap equation (\ref{B3}). For example, the one associated to the
fusing process $(1,1)+(1,1)\rightarrow (2,1)$ reads 
\begin{equation*}
S_{1b}^{12}(\theta )S_{1b}^{12}(\theta +i2\pi /3)S_{1b}^{12}(\theta +i4\pi
/3)=1\;.
\end{equation*}

\section{Conclusions}

A new set of solutions to the bootstrap equations has been constructed.
Their definition has been motivated by the semi-classical particle spectrum
of non simply-laced HSG models. The entirely new feature is the colour
structure linked to non simply-laced Lie algebras making it necessary to
combine different affine Toda models in one theory. Emphasis has been given
to show how these models can be consistently coupled to each other by
matching their different fusing structures. The universal formula obtained
for the scattering amplitude was observed to contain also the S-matrix of
simply-laced HSG models constructed earlier.

The last observation together with the chosen starting point, the
semi-classical particle spectrum of HSG theories \cite{HSGsol}, seem to
suggest that (\ref{S}) might be related to the non simply-laced HSG models,
i.e. the integrable perturbation $\Delta =\bar{\Delta}=h^{\vee }/(k+h^{\vee
})$ of the coset theory $g_{k}/u(1)^{\times \limfunc{rank}g}$ \cite{HSG}. To
draw a definite conclusion one would need to identify the conformal field
theory to which the proposed S-matrices lead in the ultraviolet limit. As
pointed out in the introduction this can be done by means of the
thermodynamic Bethe ansatz or a form factor analysis.

In case of the TBA suppose that the solutions of the TBA equations approach
a constant value at high energies. One is then in the position to proceed
analytically by solving the constant TBA equations along the lines of \cite
{TBAZ}. The effective central charge $c_{\text{eff}}$ in the UV limit would
then be given by 
\begin{equation}
c_{\text{eff}}=\frac{6}{\pi ^{2}}\sum\limits_{i=1}^{\limfunc{rank}%
g}\sum\limits_{a=1}^{k_{i}-1}\mathcal{L}\left( \frac{x_{a}^{i}}{1+x_{a}^{i}}%
\right) \quad \text{with}\quad x_{a}^{i}=\prod_{j=1}^{\limfunc{rank}%
g}\prod_{b=1}^{k_{j}-1}(1+x_{b}^{j})^{N_{ab}^{ij}}  \label{cTBA}
\end{equation}
and the exponents read 
\begin{equation}
N_{ab}^{ij}=-i\int d\theta \;\frac{d}{d\theta }\ln S_{ab}^{ij}(\theta
)=\delta _{ij}\delta _{ab}-A_{ij}t_{j}/t_{ij}\,\left(
A_{su(k_{ij})}^{-1}\right) _{\frac{t_{ij}}{t_{i}}a,\frac{t_{ij}}{t_{j}}%
b}\quad .  \label{N1}
\end{equation}
Here $\mathcal{L}(x)=\sum_{n=1}^{\infty }x^{n}/n^{2}+\ln x\ln (1-x)/2$
denotes Rogers dilogarithm \cite{Log}. This in fact would lead to a
discussion similar to the one in \cite{Resh,Kun} for RSOS models giving the
correct coset central charge of the non simply-laced HSG models. However, to
support that the constant TBA discussion is applicable here further analysis
is needed.

Alternatively, one might turn to the form factor approach. For example, all $%
n$-particle form factors of the simply-laced $su(N)$ HSG models at level $%
k=2 $ have been calculated in terms of universal determinant formulas \cite
{CFK,CF}. In the ultraviolet limit this allows to determine the conformal
operator spectrum including in particular the perturbing operator $\Delta =%
\bar{\Delta}=h^{\vee }/(k+h^{\vee })$ \cite{CF}. In this manner the
S-matrices constructed in \cite{HSGS} could be unambiguously related to the
semi-classical discussion in \cite{HSG}. One may now extend this analysis
also to the non simply-laced case with the integral formula (\ref{S1})
yielding the starting point, since from it the minimal two-particle form
factor can be immediately constructed. In contrast to the simply-laced HSG
models considered so far the non simply-laced theories proposed here will
turn out to be more complicated, since already at level $k=2$ one has to
deal with bound state poles, a feature which is absent for simply-laced Lie
algebras.\bigskip

\textbf{Acknowledgments}: I would like to thank Andreas Fring and J.Luis
Miramontes for careful reading of the manuscript and valuable comments. This
work has been started at the Freie Universit\"{a}t Berlin and has been
finished at the YITP. The financial support of the Deutsche
Forschungsgemeinschaft (Sfb288) and the National Science Foundation (Grants
DMR0073058 and PHY9605226) is gratefully acknowledged.

\end{document}